\documentclass[notitlepage, superscriptaddress
,reprint % uncomment for two-column
]{revtex4-2}
\usepackage[utf8]{inputenc}
\usepackage{bm}
\usepackage{amsmath}
\usepackage{graphicx}
\usepackage{mathrsfs}
\usepackage{color}
\usepackage{hyperref}
\hypersetup{colorlinks=true}
\usepackage{amssymb}

\usepackage{xcolor}
\usepackage[normalem]{ulem}

\begin{document}
\title{Approximate Boltzmann Distributions in Quantum Approximate Optimization}
\author{Phillip C. Lotshaw}
\email[]{lotshawpc@ornl.gov}
\affiliation{Quantum Computational Science Group, Oak Ridge National Laboratory, Oak Ridge, TN 37381, USA}
\thanks{This manuscript has been authored by UT-Battelle, LLC, under Contract No. DE-AC0500OR22725 with the U.S. Department of Energy. The United States Government retains and the publisher, by accepting the article for publication, acknowledges that the United States Government retains a non-exclusive, paid-up, irrevocable, world-wide license to publish or reproduce the published form of this manuscript, or allow others to do so, for the United States Government purposes. The Department of Energy will provide public access to these results of federally sponsored research in accordance with the DOE Public Access Plan.}

\author{George Siopsis}
\affiliation{
	Department of Physics and Astronomy, University of Tennessee, Knoxville, Tennessee  37996-1200 USA}
\author{James Ostrowski}
\affiliation{
	Department of Industrial and Systems Engineering, University of Tennessee, Knoxville, Tennessee  37996-2315 USA}
\author{Rebekah Herrman}
\affiliation{
	Department of Industrial and Systems Engineering, University of Tennessee, Knoxville, Tennessee  37996-2315 USA}
\author{Rizwanul Alam}
\affiliation{
	Department of Physics and Astronomy, University of Tennessee, Knoxville, Tennessee  37996-1200 USA}
\author{Sarah Powers}
\affiliation{Computer Science and Mathematics Division, Oak Ridge National Laboratory, Oak Ridge, TN 37381, USA}
\author{Travis S. Humble}
\affiliation{Quantum Science Center, Oak Ridge National Laboratory, Oak Ridge, TN 37381, USA}

\begin{abstract} 
Approaches to compute or estimate the output probability distributions from the quantum approximate optimization algorithm (QAOA) are needed to assess the likelihood it will obtain a quantum computational advantage. We analyze output from QAOA circuits solving 7,200 random MaxCut instances, with $n=14-23$ qubits and depth parameter $p \leq 12$, and find that the average basis state probabilities follow approximate Boltzmann distributions: The average probabilities scale exponentially with their energy (cut value), with a peak at the optimal solution. We describe the rate of exponential scaling or ``effective temperature" in terms of a series with a leading order term $T \sim C_\mathrm{min}/n\sqrt{p}$, with $C_\mathrm{min}$ the optimal solution energy.  Using this scaling we generate approximate output distributions with up to 38 qubits and find these give accurate accounts of important performance metrics in cases we can simulate exactly. 
\end{abstract}
\date{\today}
\maketitle

\section{Introduction}

Recent studies have been motivated by the prospect of obtaining a quantum computational advantage in approximately solving NP-hard combinatorial optimization problems \cite{preskill2018quantum} using the quantum approximate optimization algorithm (QAOA) \cite{farhi2014quantum}. A quantum advantage with QAOA could have important scientific impacts in diverse fields including computer science, engineering, and operations research \cite{harwood2021formulating,vikstaal2020applying,jing2022data}. Theoretical \cite{farhi2022SK,farhi2020wholegraph,basso2021quantum,marwaha2021local} and numerical \cite{lotshaw2021empirical,akshay2020reachability,akshay2022circuit,lykov2022sampling,ushijima2021multilevel} approaches have provided performance benchmarks and bounds in certain cases, typically in terms of the approximation ratio metric, which quantifies average solution quality relative to the optimal. However, the average solution quality does not directly address the success probability of QAOA in generating a single near-optimal bitstring solution to a combinatorial problem. To address this, alternative approaches have begun considering full QAOA output distributions.  A derivation showing that $p=1$ QAOA pure states have Boltzmann-distributed solution probabilities has been presented in Ref.~\cite{diez2022qaoa} based on correlations in the solution spectrums of random Ising models. Similar distributions have also been observed to describe ensemble averages of instances at $p>1$ in a ``mean-field" approach for a state based on ensemble-averaged SAT instances \cite{Hogg2000QAOA,Hogg2000QAOA2}; see also related work in Ref.~\cite{Sud2022parameter}, and on instantaneous quantum polynomial circuits in Ref.~\cite{leontica2022quantum}.  However, obtaining simplified descriptions of output from individual instances at $p>1$, and benchmarking their accuracy, have remained as important steps towards understanding performance at large depths that are relevant for quantum advantage.  
\par
We analyze output probability distributions in optimized instances of QAOA solving the MaxCut problem, for ensembles of random Erd{\H o}s-R\'enyi graphs with sizes $n \in \{14,17,20,23\}$, and with QAOA depth parameters $p\leq12$.  We find the average probability per basis state scales exponentially with costs in the MaxCut objective, similar to Boltzmann distributions in statistical mechanics, as anticipated from previous works \cite{diez2022qaoa,leontica2022quantum,Hogg2000QAOA,Hogg2000QAOA2}.  We observe systematic depth- and instance-dependent behaviors in the rate of exponential scaling or ``effective temperature", leading ultimately to a heuristic approach for generating approximate QAOA output. The estimated performance is found to accurately describe the true QAOA performance in median cases we can test exactly, while we also generate predictions for larger sizes $n \leq 38$ where we do not yet have exact results for comparison.  In total, simple Boltzmann distributions are found to give satisfactory and unified accounts of observed QAOA output distributions across diverse cases we consider.
\par
We further analyze the probability to obtain an optimal solution with QAOA.  A previous study by Ashkay {\it et.~al} has shown in simulations that a scaling $p \sim n$ suffices to obtain a fixed optimal solution probability in Max-2-SAT \cite{akshay2022circuit}, while for MaxCut the simulated probability to measure an optimal solution has been shown to decrease exponentially with $n$, as expected from the growth of the $2^n$-dimensional Hilbert space \cite{lotshaw2021empirical}. Here we analyze $p$ and $n$ dependence in the optimal solution probability for MaxCut, which is found to differ from the Max-2-SAT case.  Our analysis considers typical behavior within a specific ensemble of instances at limited sizes, and this dataset does not allow us to analyze complexity-theoretic worst-case scaling over all instances, which could vary from the typical behavior observed here.   
\par
A final component of the present work is an extension of a previous heuristic to rapidly identifying optimized variational parameters that are needed for high-performance QAOA circuits. This overcomes bottlenecks associated with variational parameter optimization, which have been suggested to limit quantum advantage in time-to-solution \cite{guerreschi2019qaoa}.  Building on Ref.~\cite{boulebnane2021predicting}, we show that a single set of parameters, taken from a previous study of the Sherrington-Kirkpatrick (SK) model at infinite size \cite{farhi2022SK}, can be rescaled and applied to the random graph ensembles we consider, with depth parameters $p\leq12$.  This extends the range of applicability of previous angle transfer heuristics, which have focused on transfer between generic or random instances at $p\leq5$ \cite{boulebnane2021predicting,lotshaw2021empirical,shaydulin2022transfer} as well as large $n$ limits or specific structured instances \cite{brandao2018fixed}, such as 3-regular graphs \cite{wurtz2021fixed,zhou2020quantum,galda2021transferability} or the SK model \cite{farhi2022SK}, in some cases with larger $p$.

\section{Quantum Approximate Optimization Algorithm}\label{sec:QAOA}

An instance of a combinatorial optimization problem is defined by a cost function  $C(\bm z)$ with an argument $\bm z = (z_1,...,z_n)$ in terms of binary variables $z_i \in \{-1,1\}$. Many scientifically relevant problems can be expressed in terms of quadratic unconstrained binary optimization for which \cite{lucas2014ising}
\begin{equation} C(\bm z) = \sum_i h_i z_i + \sum_{i,j} J_{i,j}z_iz_j\end{equation}
An optimal solution $\bm z_\mathrm{min}$ globally minimizes the cost function
\begin{equation}\label{zmin} \bm z_\mathrm{min} \in \mathrm{arg\ min}_z C(\bm z)\end{equation}
However, finding optimal solutions is often intractable due to resource requirements that scale exponentially with problem size. To overcome this problem, approximation algorithms and heuristics seek approximate solutions with reduced compute time.
\par
QAOA is a quantum heuristic designed to approximately solve combinatorial optimization problems.  The cost function $C(\bm z)$ is encoded into a quantum Ising Hamiltonian \cite{lucas2014ising}
\begin{equation} \hat C = \sum_{i,j} J_{i,j} \hat Z_i \hat Z_j + \sum_i h_i\hat  Z_i, \end{equation} 
where $\hat Z_i$ are Pauli-$Z$ operators, such that the eigenspectrum of $\hat C$ matches the set of cost function values
\begin{equation} \hat C \vert \bm z \rangle = C(\bm z)\vert \bm z \rangle . \end{equation}
Throughout this work we use hat symbols to distinguish operators, such as $\hat C$, while reserving plain symbols for functions and numbers, such as the cost function $C(\bm z)$ or a specific cost value $C$. A ground state of the operator $\hat C$ is a computational basis state $\vert \bm z_\mathrm{min}\rangle = \vert (z_1, z_2, ..., z_n)_\mathrm{min}\rangle$ that represents an optimal solution to the classical problem, while low-lying excited states represent approximate solutions. 
\par
QAOA uses a variational circuit ansatz to prepare a quantum state that is posited to return approximate ground states of $\hat C$ upon measurement in the computational basis. The QAOA ansatz uses $p$ layers that each alternate between Hamiltonian evolution under $\hat C$ and under a ``mixing" operator $\hat B = \sum_i \hat X_i$ with $\hat X_i$ the Pauli-X operator on qubit $i$,
\begin{equation} \label{QAOA} \vert \bm \gamma, \bm \beta\rangle = \prod_{l=1}^p e^{-i \beta_l \hat B}e^{-i \gamma_l \hat C}\vert \bm + \rangle, \end{equation}
where $\vert \bm + \rangle = \vert + \rangle^{\otimes n}$ is the ground state of $-\hat B$.  The $\bm \beta=(\beta_1,...,\beta_p)$ and $\bm \gamma = (\gamma_1,...,\gamma_p)$ are variational parameters chosen to minimize $\langle \hat C \rangle = \langle \bm \gamma,\bm\beta|\hat C\vert \bm \gamma,\bm\beta\rangle$; note the state notation $\vert \bm \gamma, \bm \beta\rangle$ depends implicitly on the number of algorithmic layers $p$, through $\bm \gamma$ and $\bm \beta$. When suitable parameters have been identified, then repeated preparation and measurement of $\vert \bm \gamma, \bm \beta\rangle$ yields a set of candidate solutions $\{ \bm z_\mathrm{cand}\}$ and the bitstring $\bm z^* \in \{ \bm z_\mathrm{cand}\}$ producing the smallest cost $C(\bm z)$ is taken as the final solution.  The QAOA ansatz is used because 1) in the limit $p\to \infty$ it yields the ground state and 2) the solution quality can only improve as $p$ increases \cite{farhi2014quantum}.
\par
Solution quality is often quantified through the approximation ratio
\begin{equation} r = \frac{C_\mathrm{max}-\langle \hat C \rangle}{C_\mathrm{max}-C_\mathrm{min}}\end{equation}
where $C_\mathrm{min}$ and $C_\mathrm{max}$ are the extremal values of the cost function; we computed these by evaluating $C(\bm z)$ for each $\bm z$.  Here $0 \leq r \leq 1$ quantifies the expected cost; $r=1$ signifies that QAOA prepares an optimal solution (when $\langle \hat C \rangle = C_\mathrm{min}$, in keeping with Eq.~(\ref{zmin})), while $r=0$ is the worst case.  We will further consider the probabilities for individual solutions
\begin{equation} \label{Prz} \mathrm{Pr}(\bm z) = |\langle \bm \gamma, \bm \beta\vert \bm z \rangle|^2.\end{equation}
\par
We focus on the unweighted MaxCut problem, which is a standard benchmarking problem for QAOA. An instance of MaxCut is defined with respect to a graph $G = (V,E)$ with vertex labels $V$ and edges $E = \{(i,j); i,j \in V\}$. The goal is to partition the vertices into two sets such that the number of edges with endpoints in different sets is maximized.  For a set of $n=|V|$ binary variables $z_i \in \{-1,+1\}$,  the MaxCut cost function is
\begin{equation}
    C(\bm z) = \sum_{(i,j)\in E} z_i z_j
\end{equation}
where the sum runs over the set of edges $E$ in the problem graph. The cost Hamiltonian is
\begin{equation} \hat C = \sum_{(i,j)\in E} \hat Z_i \hat Z_j.\end{equation}
We consider instance sets containing 300 random Erd{\H o}s-R\'enyi graphs at each size $n\in\{14,17,20,23\}$ \cite{hagberg2008exploring}, where in each instance each possible edge $(i,j)$ is generated with probability $0.5$.  We chose Erd{\H o}s-R\'enyi graphs rather than graphs with specified symmetries to avoid the possibility of non-generic symmetry-related behaviors.

\section{Results}\label{sec:results}

\subsection{Angle transfer}\label{sec:angle transfer}

To apply QAOA to a given problem, it is necessary to choose angle parameters $\bm \beta, \bm \gamma$ that define a QAOA circuit.  While exhaustive brute force searches yield the best possible angles, these are impractical due to the search overhead \cite{guerreschi2019qaoa, lykov2022sampling}. To remedy this problem, heuristics based on transferring angles between different problem instances have been developed and shown to achieve high performance with minimal compute time. These have included theoretical \cite{brandao2018fixed,wurtz2022counterdiabaticity} and empirical studies of transfer between graphs \cite{lotshaw2021empirical,boulebnane2021predicting,galda2021transferability,wurtz2021fixed,shaydulin2022transfer} as well as approaches for generating angles at layer $p+1$ from angles at layer $p$ \cite{zhou2020quantum,lee2022depth}. 
\par
Our approach begins with angle parameters previously devised for the Sherrington-Kirkpatric (SK) model at infinite size \cite{farhi2022SK}.  Similar angles were also identified in an analysis of large-girth graphs \cite{farhi2022girth}, with median deviations of $0.07\%$ from the SK angles and a worst-case deviation of $9\%$. Thus, the SK angles approximate optimized angles for diverse instances, at least in the large size and degree limits of Refs.~\cite{farhi2022SK,farhi2022girth}. Boulebnane and Montanaro \cite{boulebnane2021predicting} investigated an approach to rescaling these angles for applications to small Erd{\H o}s-R\'enyi and Chung-Lu graphs at $p \leq 5$, and we extend their approach to $p\leq 12$. We use relationships discussed in Refs.~\cite{boulebnane2021predicting,shaydulin2022transfer,farhi2022SK} to rescale the SK angles using the average graph degree $\overline{d} = 2|E|/n$ of a particular instance with $n$ qubits and $|E|$ edges in the problem graph.  The specific relations we use are
\begin{align} \bm \beta = \bm \beta^\mathrm{SK}, \ \ \ \ \ \bm \gamma = \frac{1}{\sqrt{\overline{d}}}\bm \gamma^\mathrm{SK}
\end{align}
where $\bm \beta^\mathrm{SK}$ and $\bm \gamma^\mathrm{SK}$ are taken from the analysis of the infinite-size SK model in Ref.~\cite{farhi2022SKarxiv}.  We then perform a gradient-based optimization from these angles, using the NLOPT implementation of the BFGS optimization algorithm \cite{nlopt,Nocedal1980bfgs,Liu1989lbfgs}, to determine optimized angles $\bm \beta^*$ and  $\bm \gamma^*$ for each individual instance. 
\par
\begin{figure}
    \centering
    \includegraphics[height=8cm,width=8cm,keepaspectratio]{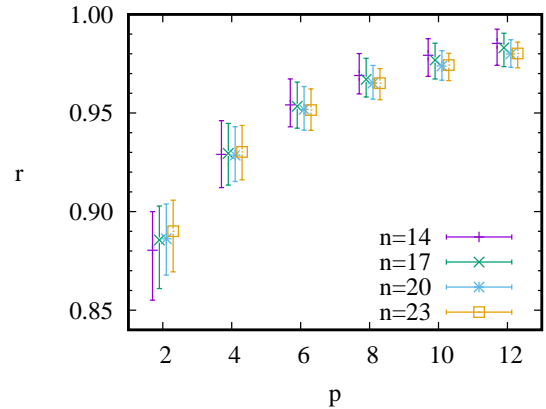}
    \caption{Approximation ratios for Erd{\H o}s-R\'enyi graphs at varying $n$ and $p=2,4,...,12$; offsets in $p$ are included for clarity.  Points show medians over the sets of 300 graphs at each $n$ and error bars show 0.1 and 0.9 quantiles.}
    \label{fig:approx ratios}
\end{figure}

Approximation ratios obtained from this procedure are shown in Fig.~\ref{fig:approx ratios}.  Points show the median approximation ratio at each $p$ for our sets of 300 graphs at each $n$, while error bars show the 0.1-0.9 quantiles over the sets of graphs.  The approximation ratios increase with $p$ and approach unity for each graph, as expected for well-optimized angles in QAOA.  This indicates qualitatively satisfactory behaviors are obtained from the rescaled and optimized SK angles.
\par
We do not have optimal angles to compare against our transfer procedure, due to the computational expense of searching for optimal angles.  Exhaustive searches in the $2p$-dimensional parameter space have used hundreds of optimizations per instance at $p\leq 3$ \cite{lotshaw2021empirical,shaydulin2022transfer} and are expected to require significantly more optimizations for our cases at $p\leq12$. As an alternative we perform a simpler comparison against a brute force search method with limited sampling of the parameter space, as described in Appendix \ref{angle transfer appendix}. We select 100 random initial angles per instance at $n=14$ at each $p$, optimize these to find a local optimum, then take the best result for each instance to compare against the transfer procedure.  We find the transferred angles obtain approximation ratios equivalent to or better than these optimized random samples.  This indicates many brute force samples are needed to outperform the much simpler transfer approach, consonant with previous works \cite{lotshaw2021empirical,shaydulin2022transfer,boulebnane2021predicting,shaydulin2022transfer,farhi2022SK,galda2021transferability,wurtz2021fixed,zhou2020quantum,lee2022depth}. The transferred angles also give similar performance at varying sizes $n$, as seen in Fig.~\ref{fig:approx ratios}, further demonstrating their practical utility. 

\subsection{QAOA Probability Distributions}

We consider probability distributions associated with the optimized QAOA states across our 300 graph ensembles at varying $n$ and $p$, beginning here with an analysis of average distributions and optimal solution probabilities, then turning to specific instances in later sections.
\par
To analyze QAOA output distributions, we compute probabilities $\mathrm{Pr}(\bm z)$ for each solution $\bm z$ and bin these to obtain the total probability to measure any solution with cost $C$,
\begin{equation} \label{PrC} \mathrm{Pr}(C) = \sum_{\bm z; C(\bm z)=C} \mathrm{Pr}(\bm z). \end{equation}
Here $C$ refers to an individual cost value, not the operator $\hat C$. In combinatorial optimization the final result is often assessed only by its cost value and probabilities for different costs are quantified by $\mathrm{Pr}(C)$.
\par
\begin{figure}
    \centering
    \includegraphics[height=18cm,width=8cm,keepaspectratio]{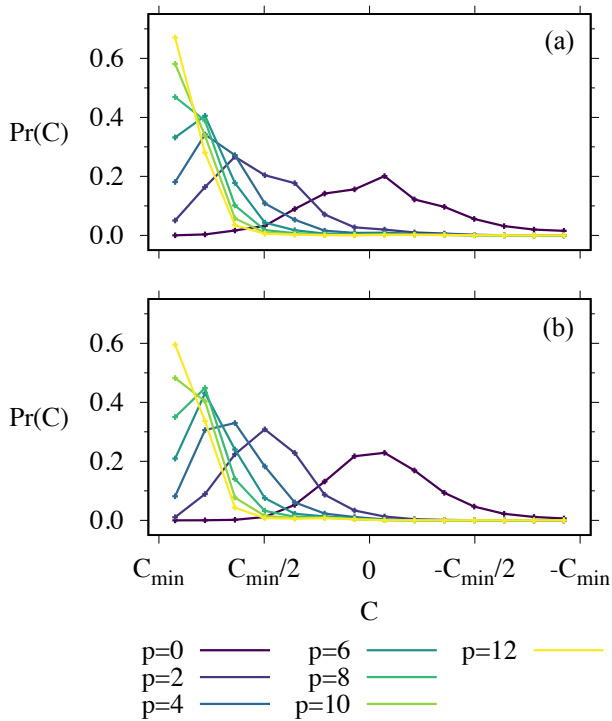}
    \caption{Average probability to measure any state with a given cost $C$, binned in widths $|C_\mathrm{min}|/7$, averaged over the 300 graph instances at (a) $n=14$ and (b) $n=23$.}
    \label{fig:cost distributions}
\end{figure}

To obtain a coarse view of how the $\mathrm{Pr}(C)$ vary with $C$ and approach the optimal $C_\mathrm{min}$ as the number of layers $p$ increases, we binned probabilities at varying costs in bins of width $|C_\mathrm{min}/7|$ for each instance. We plot the average binned probabilities in Fig.~\ref{fig:cost distributions}(a) for $n=14$ and (b) for $n=23$; we do not include standard deviation error bars as they obscure the appearance of the figure.  For simplicity we plot intervals $[C_\mathrm{min},-C_\mathrm{min}]$ though note the full distributions extend to $C_\mathrm{max}=|E|$ with very small probabilities not included in the figure. 
\par
Curves ``$p=0$" show the distributions of the initial state, corresponding to a uniform random sampling of solutions.  Here the probabilities $\mathrm{Pr}(C)$ are proportional to the densities of solutions
\begin{equation} \label{rho} \varrho(C) = \sum_{\bm z; C(\bm z) = C} 1 \end{equation}
with $\mathrm{Pr}(C)=\varrho(C)/2^n$. The average $\varrho(C)$ and $p=0$ output distributions are Gaussian-like, and the overlap with costs near the optimal $C_\mathrm{min}$ decrease as $n$ increases from $n=14$ in Fig.~\ref{fig:cost distributions}(a) to $n=23$ in Fig.~\ref{fig:cost distributions}(b). Curves for $p\geq 2$ have improved solution quality, with quasi-Gaussian average distributions centered at costs that approach $C_\mathrm{min}$ as $p$ increases.
\par
\begin{figure}
    \centering
    \includegraphics[height=8cm,width=8cm,keepaspectratio]{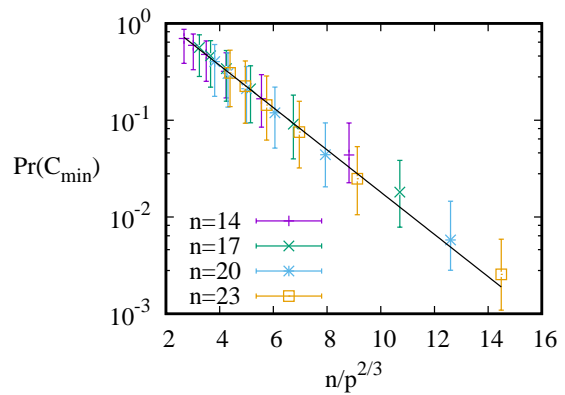}
    \caption{Probability to measure an optimal solution.  Points show medians and error bars show 0.1-0.9 quantiles across 300 graphs at each $n$ with $p\leq 12$.  The medians are fit by $a\exp(-bn/p^{2/3})$, with $a=2.75 \pm 0.09$ and $b=0.502 \pm 0.009$, with $\pm$ denoting the asymptotic standard error of the non-linear least squares fit. }
    \label{fig:PCmin scaling}
\end{figure}

The probability to obtain an optimal solution
\begin{equation} \mathrm{Pr}(C_\mathrm{min}) = \sum_{z_\mathrm{min} \in \mathrm{arg\ min}_{\bm z} C(\bm z)} \mathrm{Pr}(\bm z_\mathrm{min})\end{equation}
is shown in Fig.~\ref{fig:PCmin scaling}, with points showing the median probability at each $n$ and $p$ and error bars showing the 0.1-0.9 quantiles. We make an approximate scaling of the median $\mathrm{Pr}(C_\mathrm{min}) = ae^{-bn/p^{2/3}}$, shown by the black line, with fit parameters $a$ and $b$ in the figure caption. This indicates that to obtain a fixed probability for an optimal solution in the median case, $p$ should increase with size as $p \sim n^{3/2}$.  This scaling differs from the scaling $p \sim n$ observed in optimized instances of Max2-SAT \cite{akshay2022circuit}, which may be due to differences in the structure of the MaxCut problem, or possibly due to the choice of QAOA parameters.  The optimal solution scaling $\mathrm{Pr}(C_\mathrm{min}) \sim e^{-k_pn}$ has been observed across MaxCut instances at sizes $n \leq 9$ in previous work \cite{lotshaw2021empirical}; here we find $k_p = b/p^{2/3}$.  
\par
\begin{figure}
    \includegraphics[height=18cm,width=8cm,keepaspectratio]{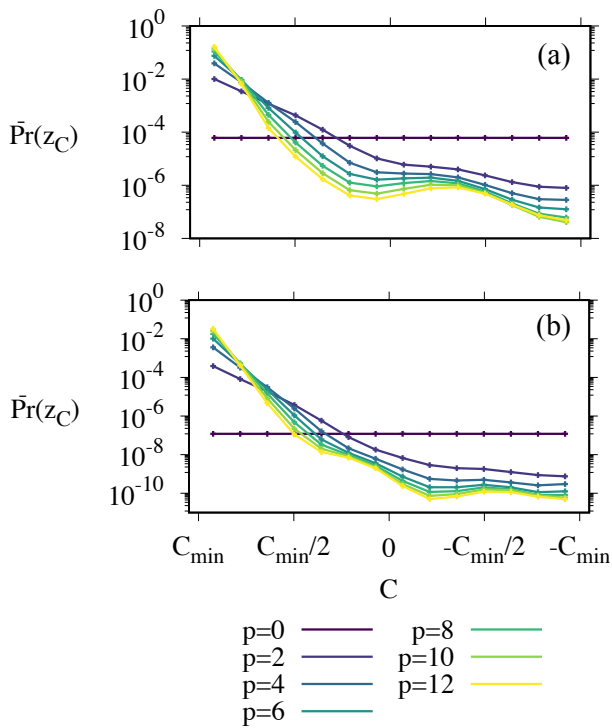}
    \caption{Average probability to measure a single basis state with cost $C$ in intervals of width $|C_\mathrm{min}|/7$, averaged over the 300 graph instances at (a) $n=14$ and (b) $n=23$.}
    \label{fig:basis probs}
\end{figure}

To obtain a more complete understanding of the sources of the distributions in Figs.~\ref{fig:cost distributions}-\ref{fig:PCmin scaling}, we consider the average probability to measure a single basis state with a given cost $C$,
\begin{equation} \label{PrzC} \overline{\mathrm{Pr}}(\bm z_C) = \frac{\mathrm{Pr}(C)}{\varrho(C)}.\end{equation}
We again consider the histogram binning procedure of previous paragraphs and plot the average $\overline{\mathrm{Pr}}(\bm z_C)$ in Fig.~\ref{fig:basis probs}(a),(b) for $n=14,23$ respectively, on a logarithmic scale.  For the $p=0$ initial state, the distribution is flat, since each individual basis state has probability $1/2^n$.  At larger $p$, the distributions increase exponentially towards $C_\mathrm{min}$, while they plateau to small values $\approx 10^{-6}-10^{-10}$ at $C\gtrsim 0$.  To a good approximation we can think of the average behavior as exponential, as solutions with $C \gtrsim 0$ contribute little overall probability.  
\par
The previous analysis gave a coarse picture of the average behavior of QAOA state distributions for random MaxCut instances, including exponential behavior of the average basis state probabilities $\overline{\mathrm{Pr}}(\bm z_C)$.  However, Figs.~\ref{fig:cost distributions} and \ref{fig:basis probs} have considered averages over many different instances, and a more detailed account of individual instances is desirable.  We address this in the next section.
\subsection{Approximate Boltzmann distributions}\label{sec:Boltzmann}

The average exponential behavior of Fig.~\ref{fig:basis probs} raises a question of whether the exponential dependence arises from averaging many instances or whether individual QAOA instances show exponential distributions. Exponential distributions were shown to approximately characterize QAOA at $p=1$ in Refs.~\cite{diez2022qaoa,leontica2022quantum}, though it is not clear if this should describe our cases with $p\geq 2$.  We consider modeling individual QAOA instances with ``Boltzmann distributions" describing the average basis state probabilities 
\begin{equation}\label{Boltzmann} \overline{\mathrm{Pr}}_\mathrm{exp}(\bm z_C) = \frac{1}{Z}e^{-C/T},\end{equation}
where the partition function $Z = \sum_C \varrho(C) e^{-C/T}$ is a normalizing constant.  $T$ controls the rate of change in basis state probabilities with varying costs.  As $T\to \infty$ the distribution becomes uniform, giving the initial QAOA state distribution ($p=0$), while the probabilities condense to the ground state as $T\to0$ ($p \to \infty$).  Eq.~(\ref{Boltzmann}) is mathematically identical to the Boltzmann distribution in statistical mechanics, though with a different conceptual foundation, as discussed in Sec.~\ref{sec:discussion}.  The total probability to obtain any solution at a given cost is then
\begin{equation} \label{PrC approx} \mathrm{Pr}_\mathrm{exp}(C) = \frac{\varrho(C)}{Z}e^{-C/T}. \end{equation}
\begin{figure}
    \centering
    \includegraphics[height=10cm,width=7cm,keepaspectratio]{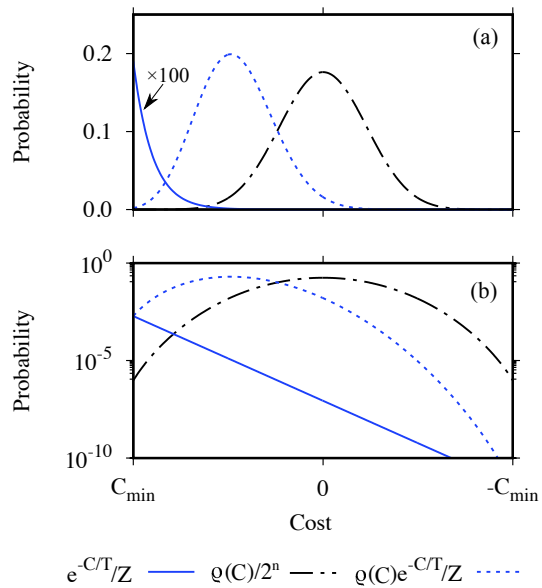}
    \caption{Schematic exponential distributions, on (a) a linear scale and (b) a log scale, see text for details. }
    \label{fig:schematic}
\end{figure}
\begin{figure*}
    \centering
    \includegraphics[height=20cm,width=\textwidth,keepaspectratio]{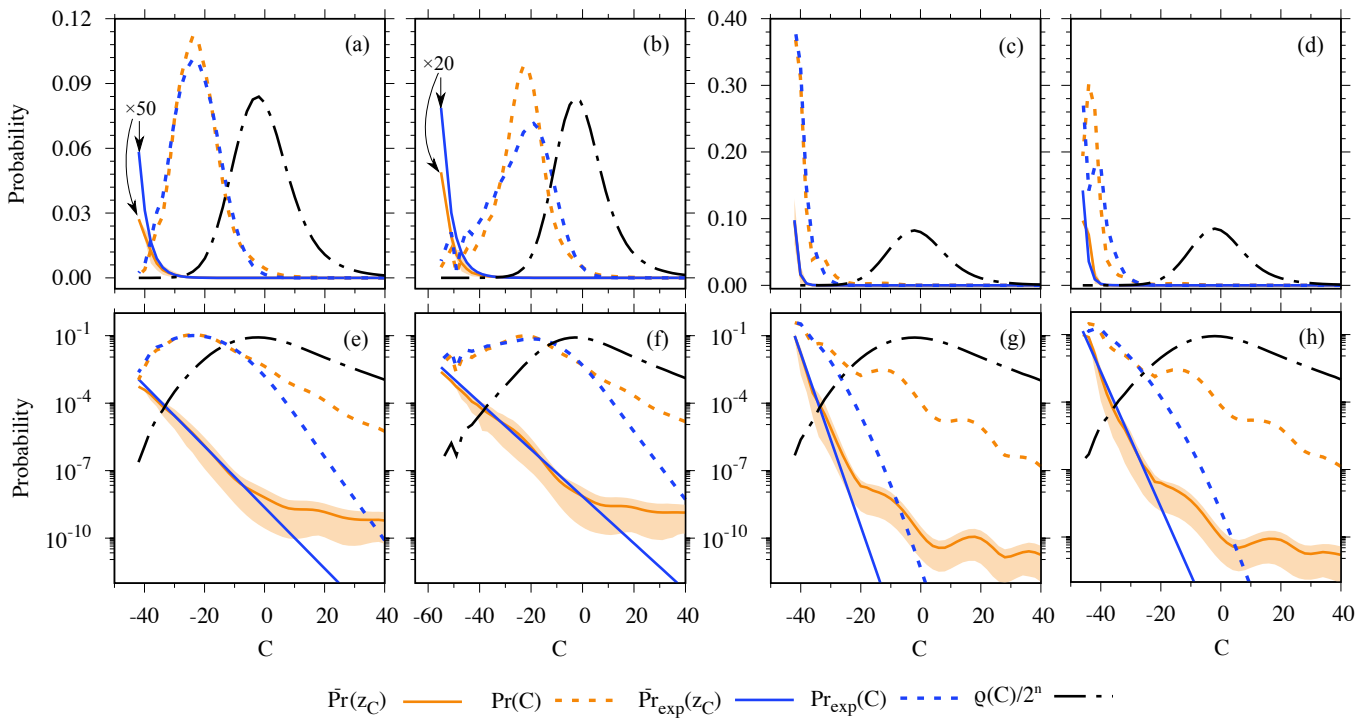}
\caption{Exact and best-fit exponential probability distributions for example instances at $n=23$ on a linear scale (top row) and logarithmic scale (bottom row). (a) and (e) show the instance with median TVD between the true and exponential distributions at $p=2$ with best-fit $T=3.20$, (b) and (f) show the worst-case TVD at $p=2$ with $T=4.16$, (c) and (g) show the median TVD at $p=12$ with $T=1.13$, and (d) and (h) show the worst TVD instance at $p=12$ with $T=1.44$. Orange and black curves are related through Eq.~(\ref{PrzC}), blue and black curves are related through Eqs.~(\ref{Boltzmann})-(\ref{PrC approx}). Shaded orange bands show the 0.1-0.9 quantiles of the $\overline{\mathrm{Pr}}(\bm z_C)$ distributions at each $C$ (too small to be visible in top panels).}
    \label{fig:distributions}
\end{figure*}
\par
Before comparing to QAOA, it may be useful to consider what to expect from the Boltzmann distribution as shown schematically in Fig.~\ref{fig:schematic} on (a) a linear scale and (b) a log scale.  The basis state probabilities  $\overline{\mathrm{Pr}}_\mathrm{exp}(\bm z_C)$ (solid blue) decay exponentially from $C_\mathrm{min}$ (Eq.~(\ref{Boltzmann})).  The number of states per cost $\varrho(C)$ (Eq.~(\ref{rho})) can be exponentially large depending on $C$; for the schematic we take $\varrho(C)$ as a binomial distribution on 20 variables with costs $\pm1$ and show a scaled version  $\varrho(C)/2^n$ (black). The total probability $\mathrm{Pr}_\mathrm{exp}(C)$ to measure any basis state with cost $C$ (dotted blue) is the product of the exponentially small $e^{-C/T}/Z$ and exponentially large $\varrho(C)$ (Eq.~(\ref{PrC approx})), with a peak at intermediate costs between zero and $C_\mathrm{min}$.  This schematically captures the expected behavior in terms of the relationships between $\overline{\mathrm{Pr}}_\mathrm{exp}(\bm z_C), \mathrm{Pr}_\mathrm{exp}(C),$ and $\varrho(C)$.  We now consider modeling QAOA probability distributions using these relations.
\par
We used non-linear least squares to fit Eq.~(\ref{PrC approx}) to exact probabilities $\mathrm{Pr}(C)$ for each graph instance at each $p$. We quantify agreement between the exponential fits $\mathrm{Pr}_\mathrm{exp}(C)$ and the true probability distributions from the wavefunction $\mathrm{Pr}(C)$ using the total variation distance $\mathrm{TVD}=\sum_C |\mathrm{Pr}_\mathrm{exp}(C)-\mathrm{Pr}(C)|/2$. If the probability distributions overlap perfectly then $\mathrm{TVD}=0$ while $\mathrm{TVD}=1$ when the distributions are disjoint.  We find significant overlaps with median $\mathrm{TVD}$ in the interval $0.034-0.095$ at each $n$ and $p$. Thus Eq.~(\ref{PrC approx}) suceeds in accounting for much of the true behavior. 
\par
We consider specific examples of true and exponential distributions at $n=23$ in Fig.~\ref{fig:distributions}.  First we consider the $p=2$ instance with the median $\mathrm{TVD}=0.056$ from among the 300 graphs at this $n$, shown in Fig.~\ref{fig:distributions}(a) on a linear scale and below in Fig.~\ref{fig:distributions}(e) on a log scale. The density $\varrho(C)$ resembles the schematic density of Fig.~\ref{fig:schematic}, but with asymmetries related to correlations in the terms $Z_iZ_j$ in the cost Hamiltonian, which are not present in the simpler schematic distribution. The exponential fit distributions (blue) show behaviors analogous to Fig.~\ref{fig:schematic} and can be understood following the discussion of that figure.  The true QAOA distributions (orange) closely resemble the exponential distributions, as expected from the previous analysis of the TVD.
\par
We now consider the worst case instance at $n=23$ and $p=2$ in Fig.~\ref{fig:distributions}(b)-(f), with $\mathrm{TVD}=0.149$. The deviations are larger, and this can be attributed to variations in the average basis state probabilities away from exponential behavior, as seen in Fig.~\ref{fig:distributions}(f); the $\overline{\mathrm{Pr}}(\bm z_C)$ are slightly above $\overline{\mathrm{Pr}}_\mathrm{exp}(\bm z_C)$ near the most probable cost $C=-23$ and slightly below the $\overline{\mathrm{Pr}}_\mathrm{exp}(\bm z_C)$ near the optimal $C_\mathrm{min}$.  The exponential distribution describes the correct qualitative behavior with semi-quantitative agreement. Furthermore, we might expect that properties of the distribution as a whole will be reproduced with smaller error, for example, probabilities that are too large or too small may tend to cancel in the approximation ratio average.  We will consider this further in the Sec.~\ref{Te perf analysis}.
\par
Fig.~\ref{fig:distributions}(c),(d),(g) and (h) are similar to the previous panels but show distributions for $n=23$ with $p=12$.  Here, the instance with the median TVD=$0.087$ is shown in (c) and (g) while the worst case instance with $\mathrm{TVD}=0.266$ is shown in (d) and (h).  This latter instance has the largest TVD from among all cases at $n=23$ (i.e., over all $n=23$ graphs at all $p$).  
\par
For the median distribution in Fig.~\ref{fig:distributions}(c), the true probabilities are concentrated on and near the optimal solution, in agreement with the exponential approximation.  For the worst case distribution in Fig.~\ref{fig:distributions}(d), the true and exponential distributions are again concentrated near the optimal solution.  However, the individual probabilities at each cost are reproduced with lower accuracy by the exponential distribution, which is peaked at $C_\mathrm{min}$ and dips at the next value $C_\mathrm{min}-2$, while the true distribution shows an opposite trend; the differences in probability near $C_\mathrm{min}$ account for the large TVD for this instance.  The exponential approximation fails to capture the individual probabilities per cost in the worst case, but it maintains a qualitative similarity to the true distribution.
\subsection{Entropy}\label{sec:entropy}

We have seen that QAOA states can be described approximately by best-fit Boltzmann distributions.  An important feature of the Boltzmann distribution is that it obtains the maximum entropy consistent with a given average energy (or cost) \cite{jaynes1957information}. Here we analyze whether the QAOA distributions match to this maximum-entropy Boltzmann expectation.  
\par
We define the Shannon entropy of the output distribution from QAOA as
\begin{equation} \label{entropy} S = -\sum_{\bm z} \mathrm{Pr}(\bm z) \log_2 \mathrm{Pr}(\bm z) \end{equation}
The entropy quantifies the uncertainty in the projective measurement distribution from QAOA \cite{NielsenChuang}; similar entropies have also been considered in analyses of quantum pure-state thermodynamics \cite{lotshaw2018micro, barnes2015quantum, lotshaw2021twobath, lotshaw2022suniv,Polkovnikov2011diagonalentropy}.  We compare QAOA state entropies against Boltzmann distribution entropies with the same $\langle \hat C \rangle$; these are different from the ``Boltzmann-fits" of the previous section, as a consistent $\langle \hat C \rangle$ is needed for a valid maximum-entropy comparison.
\par
\begin{figure}
    \centering
    \includegraphics[height=8cm,width=\columnwidth,keepaspectratio]{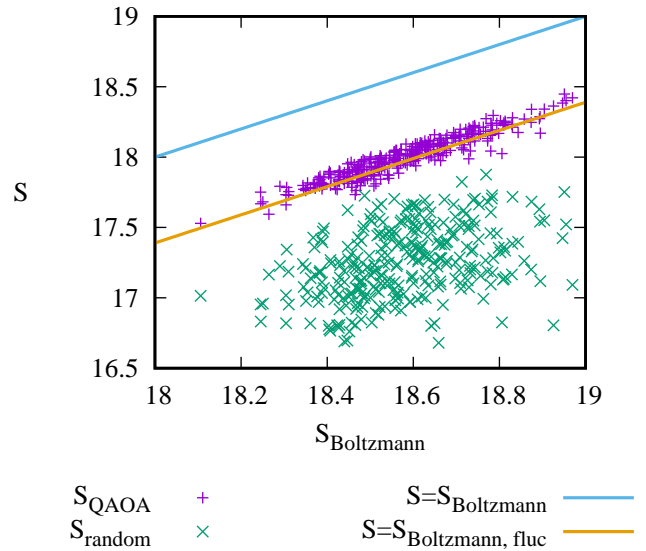}
    \caption{QAOA state entropies at $n=23$ and $p=2$, compared against entropies of random states with the same $\langle \hat C \rangle$. The QAOA state entropies are consistent with random fluctuations about a Boltzmann distribution, $S_\mathrm{QAOA} \approx S_\mathrm{Boltzmann,\ fluc}$, while the random state entropies are considerably lower.}
    \label{fig:entropy}
\end{figure}

In Fig.~\ref{fig:entropy} we compare the maximum Boltzmann entropy $S_\mathrm{Boltzmann}$ against QAOA distribution entropies $S_\mathrm{QAOA}$ at $n=23$ and $p=2$, as well as entropies of randomized states $S_\mathrm{random}$ with the same $\langle \hat C \rangle$, see Appendix \ref{entropy appendix} for details.  The QAOA states obtain sub-Boltzmann entropies (below the upper diagonal line), but these significantly exceed the entropies from the randomly generated states. Thus, the QAOA states are closer to maximum entropy than the random states, as expected from the approximate correspondence between average QAOA probabilities and the Boltzmann fits, seen previously in Fig.~\ref{fig:distributions}. 

\par
We attribute the entropy differences between the Boltzmann and QAOA distributions to two factors: First, the average QAOA state probabilities $\overline{\mathrm{Pr}}(\bm z_C)$ do not exactly follow Boltzmann distributions.  This lowers the entropy from the maximum Boltzmann value.   Second, the there are fluctuations in the individual basis states probabilities $\mathrm{Pr}(\bm z)$ about the averages $\overline{\mathrm{Pr}}(\bm z_C)$, and these fluctuations further lower the entropy. 

To better assess how fluctuations influence the entropy, we perform a final comparison against a type of state with random complex Gaussian fluctuations about a Boltzmann distribution, as described in Appendix \ref{entropy appendix}. These states can be understood analytically \cite{lotshaw2022suniv} for a straightforward comparison.  Their expected entropy is $S_\mathrm{Boltzmann,\ fluc} = S_\mathrm{Boltzmann}-(1-\gamma_\mathrm{EM})/\ln(2)$, with $\gamma_{EM}\approx 0.577$ the Euler-Mascheroni constant. Thus, for these states the fluctuations decrease the expected entropy by a constant factor $(1-\gamma_\mathrm{EM})/\ln(2)$ relative to the Boltzmann value, yielding the lower diagonal curve in Fig.~\ref{fig:entropy}, which approximately matches the QAOA entropies.  We conclude the QAOA entropy is close to what is expected for a class of states with fluctuations about average Boltzmann distributions.
\subsection{Exponential Scaling}

Here we consider the exponential scaling parameter $T$ in the best-fit Boltzmann distributions. Our goal is to devise a heuristic scaling that accounts for QAOA probability distributions across our instances.
\par

We motivate an empirical scaling formula for the temperature as follows.  We begin with the factor $\beta_T \equiv 1/T$ that appears within the Boltzmann probability factors.  To obtain a generic expression for $\beta_T$ we assume it can be expanded in powers of the number of layers $p$, about the limit $\beta_T(p=0) = 0$ which describes the initial ($p=0$) uniform superposition state.  We find an expansion in powers of $\sqrt{p/\tilde p}$ gives a good account of our results, where $\tilde p$ is an assumed constant that is characteristic of depths well above the  $p \leq 12$ we consider here.  We therefore begin with the following expression
\begin{equation} \beta_T(p/\tilde p) = \beta_T(0) + \beta_T'(0) \sqrt{\frac{p}{\tilde p}} + \beta_T''(0)\frac{p}{2\tilde p} + \ldots \end{equation} 
Using $\beta_T(0)=0$ and inverting the previous expression we find the temperature to order $p/\tilde p$
\begin{align} T & = \frac{1}{\beta_{T}'(0)}\sqrt{\frac{\tilde p}{p}}\left(1-\frac{\beta''_T(0)}{2\beta'_T(0)}\sqrt{\frac{p}{\tilde p}} \right) \nonumber\\
& = \frac{\sqrt{\tilde p}}{\beta_{T}'(0)} \frac{1}{\sqrt{p}} -\frac{\beta''_T}{2(\beta'_T(0))^2}
\end{align}
We fit this function to determine values for the unknown quantities.  We find good agreement across our 7,200 instances at varying $n$ and $p$ using the following empirical temperature relationship
\begin{equation} \label{Te} T_e = c\frac{C_\mathrm{min}}{n\sqrt{p}} + d.\end{equation}
with $c=-2.738 \pm 0.005$ and $d = -0.255 \pm 0.003$, with $\pm$ denoting the asymptotic standard error of the non-linear least-squares fit.  The factor $C_\mathrm{min}/n$ is empirical and gives an energy scale for the temperature in terms of the ground state energy per qubit; including a similar energy dependence with $d$ does not significantly effect the quality of the fit.  
\par
\begin{figure}
    \centering
    \includegraphics[height=8cm,width=8cm,keepaspectratio]{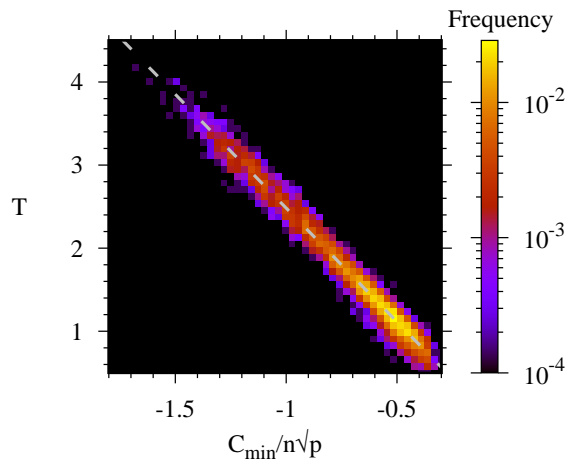}
    \caption{Two-dimensional histogram with frequencies of exponential scaling parameters $T$ compared against $C_\mathrm{min}/n\sqrt{p}$, for each of the 7,200 instances considered. The dotted line shows $T_e$ from Eq.~(\ref{Te}).}
    \label{fig:T}
\end{figure}

We examine the best-fit $T$ and the $T_e$ of (\ref{Te}) across the total set of 7,200 instances at varying $n$ and $p$. Given the large number of data points in our dataset it is difficult to visualize the results in a scatter plot.  We therefore bin instances into a two-dimensional histogram and report frequencies of occurrences, resulting in Fig.~\ref{fig:T}.   $T_e$ is shown by the dotted line in Fig.~\ref{fig:T}; it captures the typical behavior and trend, though individual instances show some deviations scattered around $T_e$.  Note the logarithmic color scale is chosen to make deviations evident, while the majority of cases are close to the heuristic $T_e$. We conclude $T_e$ gives a satisfactory account of the typical behavior within our dataset.
\par
Notably $T_e$ does not depend on the specific angles $\bm \gamma$ and $\bm \beta$ for each instance. A general characterization would need to include $\bm \gamma$ and $\bm \beta$ dependence from the QAOA states $\vert \bm \gamma,\bm\beta\rangle$. For us the $\bm \beta$ and $\bm \gamma$ for each instance are generated systematically from local optimization of the SK angles as described in Sec.~\ref{sec:angle transfer}, which evidently leads to consistent scalings $T \approx T_e$ across our instances. 
\subsection{Approximating QAOA performance metrics}\label{Te perf analysis}

We have seen that exponential average basis state probabilities are successful in reproducing approximate QAOA distributions, and further that the exponential scaling factors show systematic behavior captured by the heuristic $T_e$. This suggests typical QAOA performance can be predicted with $T_e$ and the density of solutions $\varrho(C)$, without quantum state simulations; we comment on the difficulty of computing $\varrho(C)$ in Sec.~\ref{sec:predictions}. In this section, we assess performance metric distributions generated from the $T_e$ heuristic relative to true values from the wavefunction.  We present a detailed error analysis in Appendix \ref{error analysis appendix} and refer to its main conclusions.
\begin{figure*}
    \centering
    \includegraphics[height=8cm,width=\textwidth,keepaspectratio]{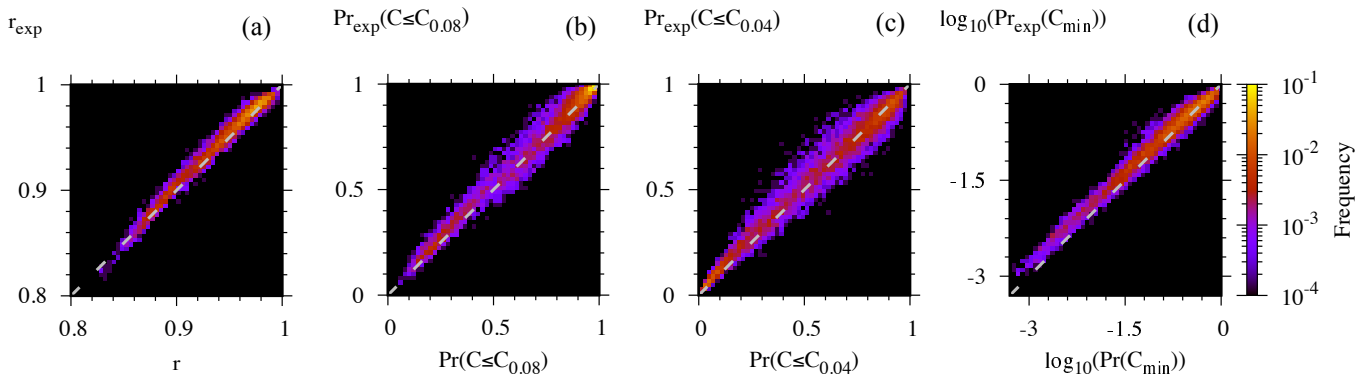}
    \caption{Two-dimensional histograms comparing exact QAOA performance metrics against results generated from the exponential distribution Eq.~(\ref{PrC approx}) with the heuristic $T_e$ of Eq.~(\ref{Te}); labels for the left-hand axes (computed with $T_e$) are shown above those axes. (a) the approximation ratio, (b) the cumulative distribution function with $C'=C_{0.08}$, (c) the cumulative distribution function with $C'=C_{0.04}$, and (d) the logarithm of the probability for an optimal solution. Dashed diagonal lines indicate where the exact and approximate results are equal.}
    \label{fig:perf metric histograms}
\end{figure*}
\par
First we consider approximation ratios $r$ and $r_\mathrm{exp}$ from the exact and approximate distributions in Fig.~\ref{fig:perf metric histograms}(a), where $r_\mathrm{exp}$ is computed using $T_e$ (Eq.~(\ref{Te})) in the exponential approximation (Eqs.~(\ref{Boltzmann})-(\ref{PrC approx})). To visualize the distributions, we use a two-dimensional histogram, showing frequencies of occurrences on a logarithmic scale. The $r_\mathrm{exp}$ slightly overestimates the true $r$, with median relative error of $0.6\%$ at each $n$ and $p$, as shown in Appendix \ref{error analysis appendix}.
\par
Next we consider the accuracy of the approximate exponential distributions for predicting the likelihood to obtain high quality results from QAOA. We consider the cumulative distribution function, which quantifies the total probability to measure any result with cost $C$ less than some threshold $C'$, 
\begin{equation} \mathrm{Pr}(C \leq C') = \sum_{C \leq C'} \mathrm{Pr}(C). \end{equation}
The cumulative distribution function is a more difficult test for our model, as it relies on predicting the total probability across a subset of costs, rather than an average over all costs as we had with $r$. These small probability subsets are less likely to contain mutually offsetting errors, so we expect the model to be less accurate than for $r$. 
\par
\begin{figure}
    \centering
    \includegraphics[height=8cm,width=8cm,keepaspectratio]{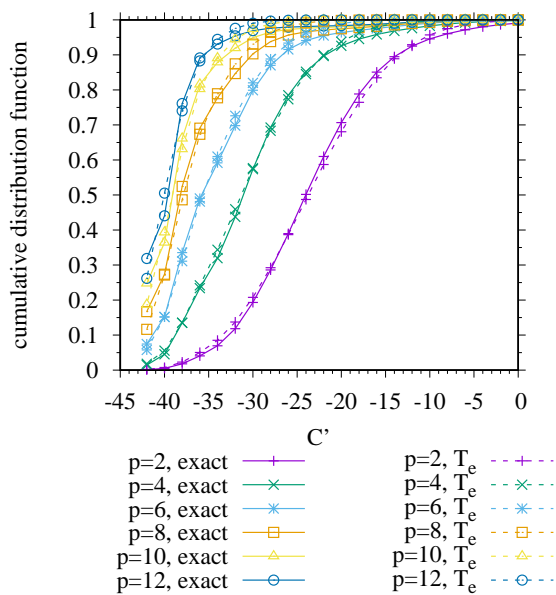}
    \caption{Cumulative distribution functions for the instance of Fig.~\ref{fig:cost distributions}(a) and (e), at varying $p$. Solid lines show the exact result $\mathrm{Pr}(C\leq C')$ while dashed lines show heuristic results $\mathrm{Pr}_\mathrm{exp}(C\leq C')$ computed with $T_e$.}
    \label{fig:cumulativedist}
\end{figure}

First we consider an example of the cumulative distribution functions at each $p$ for the example instance seen previously in Fig.~\ref{fig:cost distributions}(a),(e). Recall this had the median TVD from among all instances at $p=2$ and $n=23$, when the probabilities were approximated by the best-fit $T$. Here we consider the distributions that are generated from the heuristic $T_e$. Fig.~\ref{fig:cumulativedist} shows the cumulative distribution from the quantum state in solid lines at each $p$ as well as the distribution generated from $T_e$ in dashed lines.  The $T_e$ accurately captures the cumulative distribution for each $p$ for this instance.
\par
We assess typical performance in predicting the cumulative distribution for two values of $C'$ that focus on costs near the optimal solution. To quantify closeness to the ground state, we define costs $C_\alpha = (1-\alpha) C_\mathrm{min} + \alpha C_\mathrm{max}$.  For a given $\alpha$, any cost $C \leq C_\alpha$ is within $\alpha$ from the optimal relative to the worst case;  for example, if $\alpha = 0.08$ then any cost $C \leq C_\alpha$ is within 8\% of optimal. 
\par
Figure \ref{fig:cumulativedist} (b) and (c) show the distributions of cumulative distribution function values for $C'=C_{0.08}$ and $C'=C_{0.04}$ respectively.  Predictions from $T_e$ correlate well with true values from the wavefunction, keeping in mind that we are using a logarithmic color scale, so brighter colors are much more significant.  Deviations are greater than we had for $r$, as anticipated.  In Appendix \ref{error analysis appendix} we show that the median relative error is $\leq11\%$ at each $n$ and $p$ for $C'=C_{0.08}$, while for $C'=C_{0.04}$ the median error is $\leq35$\% at $p=2$ for each $n$, while at $p>2$ the median error is $\leq 12\%$.
\par
Finally, we consider predictions of the likelihood to measure an optimal solution $\mathrm{Pr}(C_\mathrm{min})$.  This is our most stringent test for $T_e$, as there is no chance for offsetting errors in computing a single probability. Distributions of $\log_{10}(\mathrm{Pr}(C_\mathrm{min}))$ are compared against results generated from $T_e$ in Fig.~\ref{fig:perf metric histograms}(d);  we use the logarithm to visualize the varying orders of magnitude that are present.  The predicted results show clear correlation to the true results, but with greater errors than in the previous panels of  Fig.~\ref{fig:perf metric histograms}.   At $p=2$ the median relative errors are as large as 65\%, which we attribute in part to the small values of $\mathrm{Pr}(C_\mathrm{min})$ at this $p$, which lead to large relative errors under small additive errors. The relative error decreases as $p$ increases, with median errors $\leq 29\%$ at $p>2$ across all $n$, as shown in Appendix \ref{error analysis appendix}. In the next section we further compare the scaling of $\mathrm{Pr}_\mathrm{exp}(C_\mathrm{min})$ with the scaling observed in the exact results of Fig.~\ref{fig:PCmin scaling}.
\par
The performance of $T_e$ in capturing important QAOA metrics is notable given that the $T_e$ distributions are far simpler than the exact QAOA states; they use only $T_e$ and $\varrho(C)$ to describe QAOA distributions across the 7,200 instances, as opposed to deep quantum circuit simulations. These exponential distributions give accurate accounts of the approximation ratio, with an error of $\leq1.5\%$ for 90$\%$ of cases at each $n$ and $p$, while the errors are larger for probabilities in small collections of basis states. We conclude the exponential distributions with $T_e$ capture important aspects of the true behavior. 
\subsection{Predicting performance at larger sizes}\label{sec:predictions}
We have seen that QAOA probability distributions and performance metrics can be approximated by exponential probability distributions $\mathrm{Pr}_\mathrm{exp}(C)$, using only the density of solutions $\varrho(C)$ and the scaling parameter $T_e$.  This opens the possibility of predicting performance at larger sizes $n$, using the approximate relations and without quantum state simulations.  
\par
\begin{figure}
    \centering
    \includegraphics[height=14cm,width=7cm,keepaspectratio]{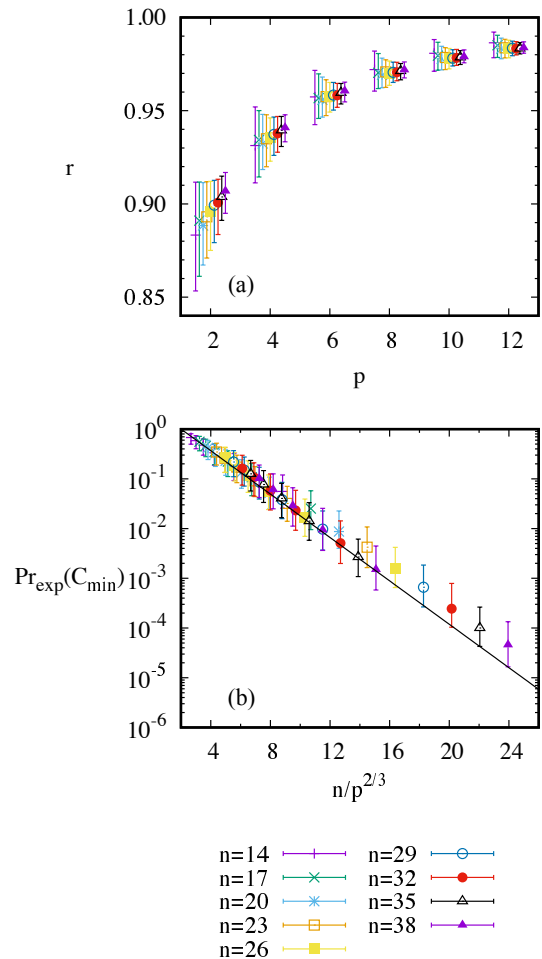}
    \caption{Predictions for (a) approximation ratios and (b) probabilities for an optimal solution, computed with Eqs.~(\ref{PrC approx}) and (\ref{Te}).  Points show medians across sets of 300 graphs at each $n$ with $p=2,4,\ldots,12$ and error bar show the 0.1-0.9 quantiles. Offsets in $p$ are included in (a) for clarity. In (b), the solid line shows the best-fit scaling from Fig.~\ref{fig:PCmin scaling}.}
    \label{fig:predictions}
\end{figure}

We computed approximate QAOA performance metrics for collections of 300 random Erd{\H o}s-R\'enyi graphs at each $n \in \{26,29,32,35,38\}$ with $p = 2,4,\ldots,12$. These use $\mathrm{Pr}_\mathrm{exp}(C)$ of Eq.~(\ref{PrC approx}) with $T_e$ from Eq.~(\ref{Te}), with densities of solutions $\varrho(C)$ computed from exhaustive enumeration of the cost values. Exactly computing $\varrho(C)$ requires time $\sim2^n$, but the time and memory requirements are small compared to what is needed for quantum state simulations.  In Fig.~\ref{fig:predictions} we plot the predicted approximation ratios and probabilities for the optimal solution at these $n$ as well as results from the previous $n\in\{14,17,20,23\}$, in all cases using the exponential distributions Eq.~(\ref{PrC approx}) with $T_e$ from Eq.~(\ref{Te}). 
\par
Approximation ratios $r$ are shown in Fig.~\ref{fig:predictions}(a).  The median $r$ increase with $n$ at small $p$, then transition to decreasing with $n$ at large $p$, and the interquantile ranges decrease with $n$, all in accord with the exact results of Fig.~\ref{fig:approx ratios}(a).  Hence, we verify that the model obtains similar results for the approximation ratio at $n\leq23$ and also predict these same behaviors extend to $n\leq 38$. 
\par
Predicted probabilities for an optimal solution are plotted in Fig.~\ref{fig:predictions}(b), with the solid line showing the best-fit relation from the exact results in Fig.~\ref{fig:PCmin scaling}.  The predicted results follow similar scaling to the fit from the exact results. It is not so clear how to derive this scaling from the exponential formalism, due to difficulties in computing the partition function $Z$ at varying $T$, but nonetheless we see the exponential distributions reproduce the known behavior.  There are deviations in the rightmost point at each $n$, corresponding to $p=2$, which may be due to larger errors in the approximate $\mathrm{Pr}_\mathrm{exp}(C_\mathrm{min})$ at this $p$ as noted in Sec.~\ref{Te perf analysis}, though it is worth noting that the exact $p=2$ medians in Fig.~\ref{fig:PCmin scaling} also slightly exceed the best fit line.  At $p\geq4$ (all points except the rightmost points at each $n$) the median predicted probabilities are close to the trend line observed in the exact results in Fig.~\ref{fig:PCmin scaling} (repeated as the black line in Fig.~\ref{fig:predictions}(b)).  Hence, we confirm the known scaling behavior of $\mathrm{Pr}(C_\mathrm{min})$ for $n\leq 23$ and predict similar scaling extends to $n\leq 38$.  The predictions depicted in Fig.~\ref{fig:predictions} can be tested in future work, through simulations or direct sampling from a high-fidelity quantum device.
\section{Discussion}\label{sec:discussion}

 The systematic behavior in the approximate QAOA Boltzmann probabilities is surprising. One reason is that the Boltzmann factor $e^{-C/T}$ reflects global structure in the output distribution, in terms of the variable $C=C(\bm z)$ that depends on all qubits, while QAOA is usually understood as a local algorithm. For example, it is well known that QAOA dynamics generate local correlations in terms of $p$-dependent ``subgraphs", and this has often been taken as a starting point for analyzing QAOA \cite{farhi2014quantum,wurtz2021maxcut,farhi2022girth,marwaha2021local,brandao2018fixed}.  The current work and Refs.~\cite{diez2022qaoa,leontica2022quantum,Hogg2000QAOA,Hogg2000QAOA2,Sud2022parameter} have assessed the opposite extreme, in terms of global structure.  Exploring the emergence of global structure from local dynamics appears likely to yield new insights into structure and behaviors of QAOA, potentially leading to new ways to predict performance similar to Sec.~\ref{sec:predictions}, or to new approaches for understanding prospects for quantum advantage.
\par
There is an interesting connection between the present work and statistical mechanics, where the Boltzmann distribution plays a prominent role in describing microscopic state probabilities in the presence of a heat bath at temperature $T$.  An important difference here is that the QAOA distribution arises from pure state dynamics alone, without any exterior system or bath.  The QAOA states resemble approximate ``thermal pure quantum states" in the terminology of Ref.~\cite{Sugiura2012thermalpure}, which proposed such states for fundamental investigations of thermodynamic behavior.
\par
The framework of statistical mechanics opens interesting possibilities for future analysis of QAOA. One interesting aspect of physical Boltzmann distributions is that they concentrate in energy, with fluctuations $\Delta E/E$ that tend to zero as the size of the system increases \cite{plischke1994equilibrium}.  This is the statistical mechanical explanation for why macroscopic systems have an effectively constant energy at thermal equilibrium. It could be interesting to apply similar reasoning to QAOA, to argue for $p$-independent concentration at large $n$, which might provide an alternative route to establishing concentration results beyond small $p$ analyses \cite{farhi2014quantum} or detailed considerations of specific instances \cite{lykov2022sampling, brandao2018fixed}.  To establish such an analogy it appears necessary to derive exponential distributions from the QAOA ansatz, to better assess their prevalence and limitations. This is an interesting topic for future work and may build on initial efforts of Refs.~\cite{diez2022qaoa,Hogg2000QAOA,Sud2022parameter}.
\par
It is important to clarify that we have analyzed a dataset at small $n$ and $p$ and that different behavior might be observed at large $n$ and $p$.  For worst-case instances it is expected that exponential time will be needed to reach exact solutions, which could arise for example if the number of algorithmic layers or the time for angle optimization scales exponentially. At the same time, we find systematic behavior across small typical instances, and it is certainly worth testing how well this extends to larger typical instances as well as comparing against worst-case behavior. 
\par
\begin{figure}[h!]
    \centering\includegraphics[height=14cm,width=\columnwidth,keepaspectratio]{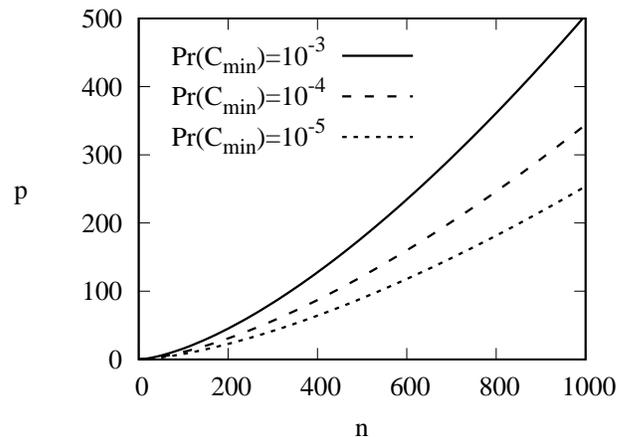}
    \caption{Estimated number of layers $p$ to obtain a fixed optimal solution probability $\mathrm{Pr}(C_\mathrm{min})$, based on the empirical scaling in Figs.~\ref{fig:PCmin scaling},\ref{fig:predictions}.}
    \label{fig:optimal scaling}
\end{figure}

Although we cannot know the true large size scaling for QAOA given our dataset, it may nonetheless be useful to assess what would be expected if the scaling we observed were to extend to larger sizes. How many layers would be needed to identify an optimal solution at larger $n$ with a modestly high probability? From our scaling we estimate that to obtain a fixed optimal solution probability, the estimated $p$ at larger $n$ entails resources that may be very difficult to achieve on near-term hardware, unless the true large-size scaling is actually much more forgiving than what we have observed at small sizes. Fig.~\ref{fig:optimal scaling} shows the size-dependent number of layers for varying optimal solution probabilities $\mathrm{Pr}(C_\mathrm{min})$ (or varying expected numbers of measurements $1/\mathrm{Pr}(C_\mathrm{min})$), based on the scaling in Figs.~\ref{fig:PCmin scaling},\ref{fig:predictions}.  We estimate hundreds of layers are expected at sizes of $n\sim 10^3-10^4$, in the absence of noise. Implementing these deep circuits with high fidelity would require very low noise levels. There may also be non-linear behaviors that were not detected in our dataset but that become dominant at larger $n$ or $p$, which may further limit performance. Overall these extrapolations suggest demanding resource requirements for identifying optimal solutions with QAOA at scale, unless the true large size scaling is more forgiving than what is suggested by our fitted small-size trend.
\par
Finally, it is worth noting that generating samples from exponential distributions is of independent interest for a variety of applications.  These include understanding the thermodynamic properties of materials \cite{Sugiura2012thermalpure}, optimization with simulated annealing \cite{Kirkpatrick1983simulatedannealing}, and machine learning with ``Boltzmann machines" \cite{Montufar2018RBM}. If QAOA can quickly generate exponentially distributed samples then it might find use in new quantum-classical algorithms. 

\section*{Data availability}

Simulation results are available at:  https://code.ornl.gov/5ci/approximate-boltzmann-distributions-in-quantum-approximate-optimization

\begin{acknowledgements}
P.C.L.~thanks Ryan Bennink for interesting discussions of combinatorial optimization, feedback, and encouragement. This work was supported by DARPA ONISQ program under award W911NF-20-2-0051. J. Ostrowski acknowledges the Air Force Office of Scientific Research award, AF-FA9550-19-1-0147. G.\  Siopsis  acknowledges the Army Research Office award W911NF-19-1-0397. J.\ Ostrowski and G.\ Siopsis acknowledge the National Science Foundation award OMA-1937008. This research used resources of the Compute and Data Environment for Science (CADES) at the Oak Ridge National Laboratory, which is supported by the Office of Science of the U.S. Department of Energy under Contract No. DE-AC05-00OR22725.
\end{acknowledgements}

\bibliography{references}

\appendix

\section{Angle transfer compared to brute force search} \label{angle transfer appendix}

To assess performance of our angle transfer approach from Sec.~\ref{sec:angle transfer}, we compared against a more standard approach based on a brute force search with a modest amount of sampling.  We generated 100 random initial angles for each instance at $n=14$ and $p=2,4,...,12$.  We optimized each random set of angles with the BFGS algorithm to identify a nearby local optimum, then selected the angles $\bm \beta^*$ and $\bm \gamma^*$ that obtained the largest approximation ratio for each instance.  We denote these approximation ratios $r_\mathrm{bf}$ and compare them against our transfer approach of Sec.~\ref{sec:angle transfer}, denoted here as $r_\mathrm{tr}$. Similar brute force approaches have been considered in previous publications, which have concluded that angle transfer achieves similar performance to brute force optimization at $p\leq 5$ \cite{lotshaw2021empirical,shaydulin2022transfer,boulebnane2021predicting}. 
\par
\begin{figure}
    \centering
    \includegraphics[height=8cm,width=8cm,keepaspectratio,trim={2cm 0 0 0},clip]{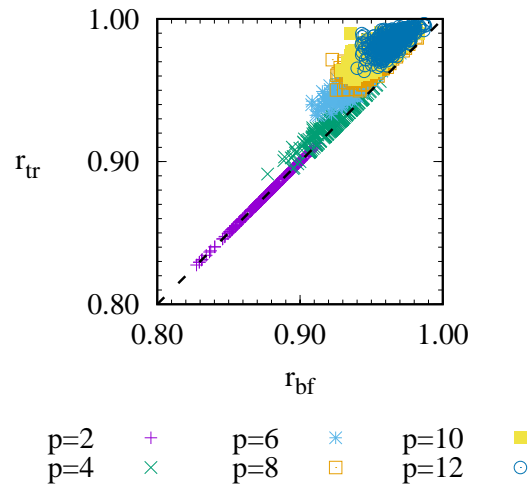}
    \caption{Approximation ratios from our angle transfer procedure $r_\mathrm{tr}$ are equivalent or larger than approximation ratios $r_\mathrm{bf}$ from a brute force approach that optimizes 100 random initial angles for each instance to identify a best set of parameters. The dashed diagonal line shows where $r_\mathrm{tr} = r_\mathrm{bf}$.}
    \label{fig:transfer bf}
\end{figure}

Figure \ref{fig:transfer bf} compares the approximation ratios obtained from the two approaches.  At $p=2$, the results are equivalent for all cases, while at $p\geq 4$ the transfer approach performs equivalently or better in all cases, $r_\mathrm{tr} \geq r_\mathrm{bf}$.  This confirms that our angle selection procedure is competitive with a more standard brute force approach, consonant with previous analyses at $p\leq 5$. 
\par
A more exhaustive analysis of performance of our procedure would require greater numbers of random initial samples for the brute force case.  When the number of samples is sufficiently large, this should identify the global optimal angles and yield results that cannot be worse than the transfer approach.  However, this may require a very large number of samples, as the size of the $\bm \gamma, \bm \beta$ parameter space increases exponentially with $p$. This makes the brute force searches impractical. By contrast, the angle selection procedure we have used here is much more scalable, and achieves impressive results in Fig.~\ref{fig:transfer bf} and Fig.~\ref{fig:approx ratios}. 

\section{Entropy analysis}\label{entropy appendix}

Here we present details of how each entropy is computed in Sec.~III.D. The entropy of the QAOA states $S_\mathrm{QAOA}$ is evaluated directly from the wavefunction probabilities $\mathrm{Pr}(\bm z)$ following Eqs.~(7) and (17). The Boltzmann entropy is defined using the Boltzmann distribution $\mathrm{Pr}_\mathrm{Boltzmann}(\bm z) = e^{-C(\bm z)/T_B}/Z$ with the same expected cost $\langle \hat C \rangle$ as a given QAOA instance, and with $Z = \sum_{\bm z} e^{-C(\bm z)/T_B}$.  We compute the Boltzmann distribution and its entropy following the approach of minimizing the ``$\Gamma$ potential" of Ref.~\cite{mead1984maximum} to identify a $T_B$ that satisfies the standard statistical mechanical relation
\begin{equation} - \frac{\partial \ln Z}{\partial(1/T_B)} = \langle \hat C \rangle . \end{equation}
The Boltzmann entropy is then computed using the standard relation (expressed in units of $\log_2$) 
\begin{equation} S_\mathrm{Boltzmann} = \frac{1}{\ln(2)} \left(\frac{\langle\hat C \rangle}{T_B} + \ln Z\right) \end{equation}
\par
The entropy $S_\mathrm{Boltzmann,\ fluc}$ is devised following the approach of Ref.~\cite{lotshaw2022suniv}.  We consider a quantum state with random independent complex Gaussian fluctuations about the Boltzmann distribution
\begin{equation} \vert \psi\rangle = \sum_{\bm z} \tilde g_{\bm z} \sqrt{\frac{e^{-C(\bm z)/T_B}}{Z}}\vert \bm z \rangle  \end{equation}
Here $\tilde g_{\bm z} = (g_{\bm z} + ig_{\bm z}')/\sqrt{2}$, where each of $g_{\bm z}$ and $g_{\bm z}'$ are taken as random independent Gaussian variates, with probability density $\mathrm{Pr}(g_{\bm z})=e^{-g_{\bm z}^2/2}/\sqrt{2\pi}$ and similarly for $g_{\bm z}'$.  The expected entropy is then approximated by taking expectation values for the terms containing $\tilde g_{\bm z}$, which yields the correction factor $- (1-\gamma_{EM})/\ln(2)$ in addition to the Boltzmann entropy \cite{lotshaw2022suniv}.
\par
We generate the random states with the same $\langle \hat C \rangle$ as the QAOA states and compute their entropies as follows.  We consider a set $F$ of all the probability assignments consistent with the constraints
\begin{align}
\sum_C \mathrm{Pr}(C) C = \langle \hat C\rangle \nonumber\\
\sum_C \mathrm{Pr}(C) = 1\nonumber\\
\mathrm{Pr}(C) \geq 0 &\  \forall C
\end{align}
To generate a randomized probability set $\{\mathrm{Pr}(C)\} \in F$, we first generate a uniformly random set of probabilities $\{\mathrm{Pr}'(C)\}$. We then find the closest point $\{\mathrm{Pr}(C)\} \in F$ with respect to the $L_2$ metric, by minimizing
\begin{equation} \min_{\{\mathrm{Pr}(C)\} \in F} \sum_C (\mathrm{Pr}'(C) - \mathrm{Pr}(C))^2 \end{equation}
\begin{figure*}
    \centering
    \includegraphics[height=50cm,width=\textwidth,keepaspectratio]{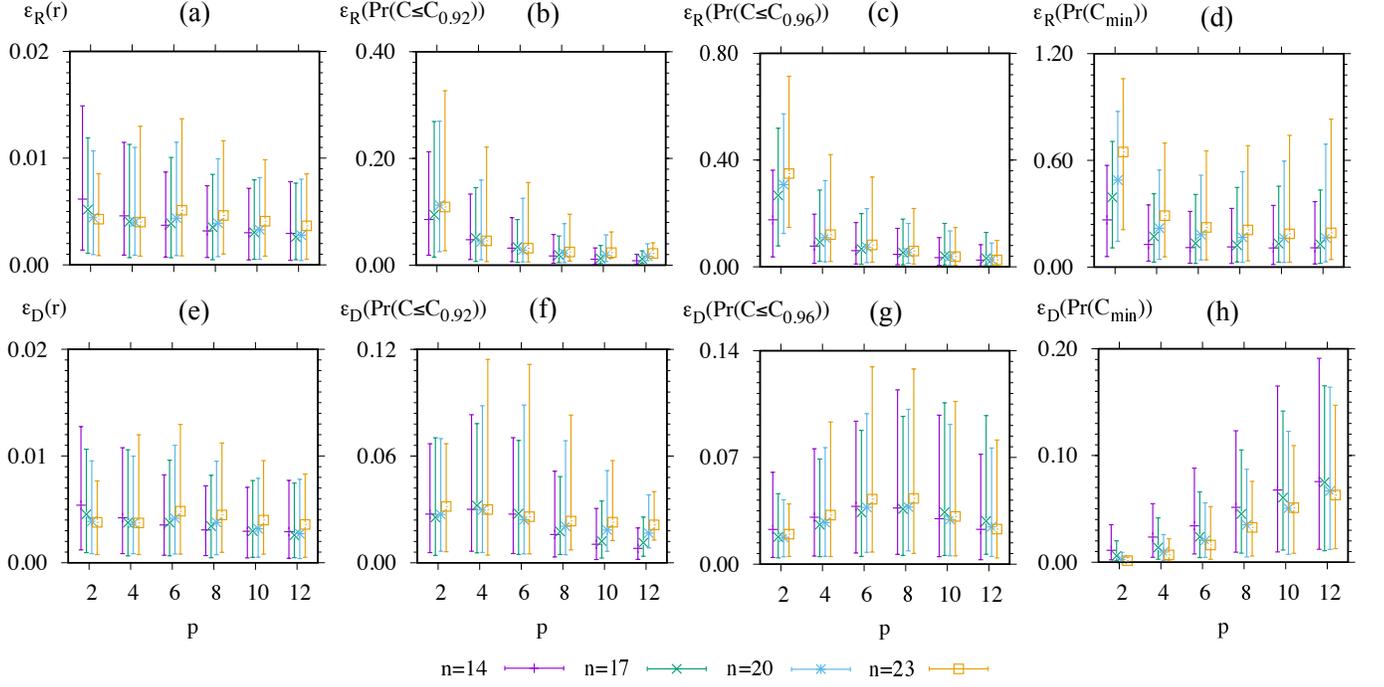}
    \caption{Errors in approximating various QAOA performance metrics with exponential distributions using the empirical scaling factor $T_e$.  The top rows shows relative errors, $\epsilon_R$ of Eq.~(\ref{epsilonR}), while the bottom row shows difference errors, $\epsilon_D$ of Eq.~(\ref{epsilonD}). Each column shows different performance metrics: (a),(e) the approximation ratio $r$, (b),(f) the cumulative distribution function $\mathrm{Pr}(C\leq C_{0.92})$, (c,g) the cumulative distribution function $\mathrm{Pr}(C\leq C_{0.96})$, and (d),(h) the probability of to measure an optimal solution. Points are medians over the sets of 300 graphs at each $n$ and error bars show the 0.1-0.9 quantiles.}
    \label{fig:r_P_stats}
\end{figure*}
To perform this minimization we use the software package {\it Gurobi}.  The result is a randomized probability assignment $\{\mathrm{Pr}(C)\}$ consistent with the QAOA expected cost $\langle \hat C\rangle$. Finally, for each cost $C$ we generate $\varrho(C)$ different basis state probabilities $\mathrm{Pr}(\bm z)$ uniformly at random and normalize these probabilities so they sum to $\mathrm{Pr}(C)$.  We then compute the entropy Eq.~(17) using these randomized $\mathrm{Pr}(\bm z)$. The resulting entropy corresponds to a randomized state with the same $\langle \hat C \rangle$ as the QAOA state. 

\section{Error analysis of approximate QAOA performance}\label{error analysis appendix}
We define relative and absolute errors to compare the simple approximate QAOA distributions with $T_e$, from Eqs.~(16) and (20), to the true QAOA distributions in Eq.~(11).  We define the absolute relative error as
\begin{equation}\label{epsilonR} \epsilon_R(Q) = |1-Q_\mathrm{exp}(T_e)/Q|,\end{equation}
where $Q$ is the quantity of interest, for example, the approximation ratio $Q=r$ or probability of the ground state $Q=\mathrm{Pr}(C_\mathrm{min})$.  Similarly we define the absolute difference error as
\begin{equation}\label{epsilonD} \epsilon_D(Q) = |Q_\mathrm{exp}(T_e)-Q|.\end{equation}
In each case $\epsilon=0$ signifies zero error, while the amount of error increases with $\epsilon$.  In Fig.~\ref{fig:r_P_stats} we analyze errors for each of the quantities considered previously in Fig.~9 of Sec.~III.F. We consider median errors, denoted by points in Fig.~\ref{fig:r_P_stats}, with error bars showing the 0.1-0.9 quantiles.  The top row shows the relative absolute error while the bottom row shows the absolute difference error.
\par
In Fig.~\ref{fig:r_P_stats}(a) and (e) we plot errors $\epsilon_R(r)$ and $\epsilon_D(r)$ respectively. The median error is $\leq0.6\%$  at each $n$ and $p$, while $90\%$ of cases at each $n$ and $p$ have relative errors $\leq1.5\%$, as shown by the error bars. The worst-case error is $2.9\%$.  We conclude that $T_e$ gives very accurate predictions for the approximation ratio across the 7,200 instances we consider. 
\par
In Fig.~\ref{fig:r_P_stats}(b) and (f) we assess error in predicting the cumulative distribution function $\mathrm{Pr}(C\leq C_{0.08})$.  The relative errors are largest at $p=2$, and decrease as $p$ and $\mathrm{Pr}(C\leq C_{0.08})$ increase.  The absolute difference error in Fig.~\ref{fig:r_P_stats}(f) shows only minor variations across $n$ and $p$, consistent with larger relative errors at small $p$, when $\mathrm{Pr}(C\leq C_{0.08})$ is small.  The median relative error in $\mathrm{Pr}(C\leq C_{0.08})$ from the approximation with $T_e$ is $\leq11\%$ for median cases, and $\leq33\%$ in 90\% of cases, as seen by the 0.9 quantile error bars. Thus, for the majority of instances at each $n$ and $p$, we obtain an accurate prediction of $\mathrm{Pr}(C\leq C_{0.08})$ to within a factor of 1.33 or better. Larger errors are observed for the cumulative distribution function $\mathrm{Pr}(C\leq C_{0.04})$ in Fig.~\ref{fig:r_P_stats}(c) and (g), as anticipated from the previous analysis of Fig.~9.  At  $p=2$ the median relative error is $\leq35\%$, when  $\mathrm{Pr}(C\leq C_{0.04})$ is small. At $p>2$ the median relative error decreases to $\leq12\%$ across all $n$ and $p$.
\par
Finally, we consider predicted probabilities to observe an optimal solution $\mathrm{Pr}(C_\mathrm{min})$.  Here the absolute error is close to zero at $p=2$, when the probability in optimal solutions is small. These small absolute difference errors result in median relative errors $\leq65\%$. The difference errors increase with $p$, as more population enters the optimal solution states, while the relative errors decrease for a similar reason. The median errors are $\leq29\%$ at $p>2$, indicating a better approximation for these $p$.
\par
Overall,  the median errors tend to increase with $n$, as do the spread of the errors, as reflected in the 0.1-0.9 quantile error bars. Nonetheless, the increase in median relative error with $n$ is fairly modest for most cases, especially at $p>2$, which indicates versatility of the approach in predicting median case behavior over our graph ensembles.

\section{Supplemental Figures}

\begin{figure}
    \centering
    \includegraphics[width=7cm,height=23cm,keepaspectratio]{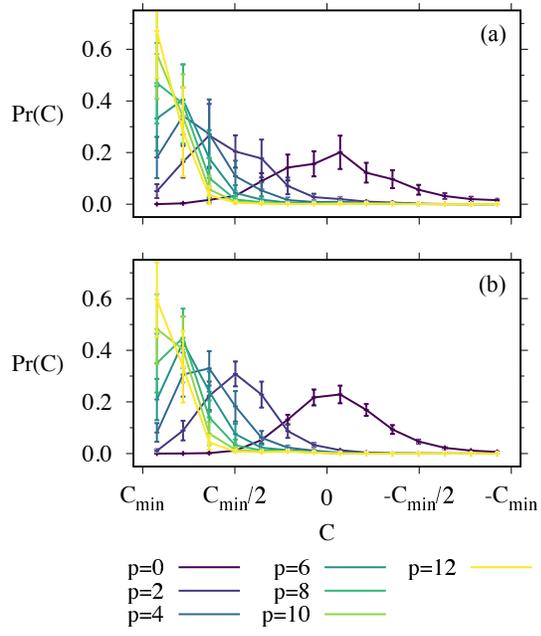}
    \caption{Same as Fig.~2 but with standard deviation error bars.}
    \label{fig:enter-label}
\end{figure}

In Fig.~\ref{fig:cost distributions} we did not include error bars because they crowded the curves in the figure.  Here in Fig.~\ref{fig:enter-label} we present a version with error bars to gauge the size of deviations among varying instances.

\end{document}